\documentclass[letterpaper,titlepage,10pt]{article}
\usepackage{hyperref}
\usepackage{amssymb,amsmath,amsfonts}
\usepackage{epsfig}

\def\clock{{\count0=\time
           \divide\count0 60
           \ifnum\count0<10 0\fi\the\count0
           \multiply\count0 -60 \advance\count0 \time
           :\ifnum\count0<10 0\fi \the\count0
         }}
\newcommand{\timestamp}{{\small\vbox{\hbox{\tt\jobname.tex}
\hbox{\the\day/\the\month/\the\year, \clock}}}}

\newcommand{\void}[1]{}

\newcommand{\Z}{\mathbb{Z}}
\newcommand{\C}{\mathbb{C}}

\makeatletter
\def\mr@ignsp#1 {\ifx\:#1\@empty\else #1\expandafter\mr@ignsp\fi}%
\newcommand{\multiref}[1]{\begingroup
\xdef\mr@no@sparg{\expandafter\mr@ignsp#1 \: }%
\def\mr@comma{}%
\@for\mr@refs:=\mr@no@sparg\do{\mr@comma\def\mr@comma{,}\ref{\mr@refs}}%
\endgroup}
\makeatother

\newcommand{\hypref}[2]{\ifx\href\asklfhas #2\else\href{#1}{#2}\fi}

\renewcommand{\eqref}[1]{(\multiref{#1})}

\def\[{\begin{equation}}
\def\]{\end{equation}}
\newcommand{\be}{\begin{eqnarray}}
\newcommand{\ee}{\end{eqnarray}}

\def\bea{\begin{eqnarray}}
\def\eea{\end{eqnarray}}

\usepackage{color}

\begin{document}

\begin{titlepage}

\rightline{\vbox{\small\hbox{\tt NITS-PHY-2011003} }}
\vskip 1.8 cm

\centerline{\LARGE \bf New symmetries of the chiral Potts model}
\vskip 0.2cm
\vskip 1.5cm

\centerline{\large {\bf Jens Fjelstad$\,^{1}$} and {\bf Teresia M\aa{}nsson$\,^{2}$},}

\vskip 0.5cm

\begin{center}
\sl $^1$Department of Physics\\
Nanjing University\\
22 Hankou Road, Nanjing,
210093 China 
\vskip 0.4cm
\sl $^2$ Department of Theoretical Physics, School of Engineering Sciences\\
Royal Institute of Technology(KTH) \\
Roslagstullsbacken 21,
SE-106 91 Stockholm
\sl   \\
\end{center}
\vskip 0.5cm

\centerline{\small\tt 
jens.fjelstad@gmail.com, teresiam@kth.se}

\vskip 1.5cm 
\centerline{\bf Abstract} 
\vskip 0.2cm 
\noindent 
	In this paper a hitherto unknown spectrum generating algebra, consisting of two coupled Temperley-Lieb algebras, is found in the three-state chiral Potts model.
From this we can construct new Onsager integrable
models. One realisation is in terms of a staggered isotropic XY spin chain.
Further we investigate the importance of the algebra for the existence of mutually commuting 	conserved charges.
This leads us to a natural generalisation of the boost-operator, which generates the charges.

\end{titlepage}

\tableofcontents


\section{Introduction}
\label{sec:Intro}

The Temperley-Lieb (TL) algebra \cite{Temperley:1971} plays an important role for the integrability of some basic statistical models, see for instance \cite{Baxter:1982, Martin:1991} and references therein.
 Solvability of a model relies on the existence of a large enough symmetry, and in certain models the special properties of the TL algebra
	(playing the role of a spectrum generating algebra)
 guarantees the existence of a boost (or ladder) operator from which all mutually commuting conserved (local) charges can be generated. Another, partially overlapping, set of models are integrable due to having a Hamiltonian of a special form satisfying the so called Dolan-Grady condition \cite{DolanGrady}. 
These models exhibit \cite{Perk:1987} a symmetry generated by the infinite dimensional so-called Onsager algebra \cite{Onsager:1943jn}, and those models which are integrable in both senses are often called superintegrable \cite{ITP-SB-88-16}.

In this paper we concern ourselves with the three state chiral Potts chain \cite{Howes:1983mk}, which for special values of certain parameters is known to be superintegrable \cite{ITP-SB-88-16}. Except for a degenerate point in the parameter space, coinciding with the conventional three state Potts chain, a boost operator generating conserved charges is not known, and the TL algebra is not known to play any role. We show in section \ref{sec:cp} that the chiral Potts chain Hamiltonian can be expressed as a representation of an element in an associative algebra generated from two copies of a TL algebra. We then abstract a one--parameter class of algebras$, \mathcal{A}_n(\alpha)$, generalising the one found in the chiral Potts chain. The complicated form of the relations between the two copies of the TL algebra results in a structure which is quite difficult to analyse in general, e.g. we were so far not able to determine the dimension of this algebra for a chain of fixed length.
Section \ref{sec:integrab} is devoted to an investigation of general nearest neighbour Hamiltonians, expressed in terms of representations of $\mathcal{A}_n(\alpha)$, with respect to integrability. For a Hamiltonian of this form satisfying a condition generalising the chiral Potts integrability condition, we find a derivation of $\mathcal{A}_n(\alpha)$ that we adopt as a candidate for (the derivation w.r.t.) a boost operator. It is shown explicitly that the first of the recursively defined charges is conserved, and it automatically follows that also the second is. A computer calculation furthermore confirms that these first two charges mutually commute. No example of this type is known where only the first charges are conserved, and we conjecture that all charges generated from this derivation are conserved and mutually commute (see also Conjecture 2 in \cite{Grabowski:1994rb}).
In section \ref{sec:Onsager} we focus on Onsager integrability and the Dolan-Grady condition. The relations of $\mathcal{A}_n(\alpha)$ is shown to imply the Dolan-Grady condition for a set of Hamiltonians generalising the superintegrable Hamiltonians in the chiral Potts chain.
Section \ref{sec:realisations} is devoted to realisations of the coupled TL algebras in other models. We find that the staggered XXZ Heisenberg, and the staggered isotropic XY Heisenberg models can be expressed in terms of representations of these algebras. In particular it follows that the latter model is superintegrable for a certain choice of parameters. We furthermore find a class of models generalising the $SU(N)$ spin chains of \cite{Affleck} which exhibit a closely related 
	spectrum generating algebra
 also defined by two coupled TL algebras, but not satisfying all relations of $\mathcal{A}_n(\alpha)$. Generically, however, these generalisations break the $SU(N)$ symmetry down to a Cartan torus $U(1)^{N-1}$.
 In Section \ref{sec:disc} we conclude with brief discussions of open questions. Finally, we include four appendices containing technical details; Appendix \ref{app:ids} contains relations which are useful in dealing with the algebras defined in Section \ref{sec:cp}, Appendices \ref{app:boost} and \ref{app:onsager} contain details of the calculations in Sections \ref{sec:integrab} and \ref{sec:Onsager} respectively, and Appendix \ref{small} contains a $9$-dimensional representation of $\mathcal{A}_2(5/2)$ which, together with the chiral Potts representation, has been used to double check a majority of the presented results.

\section{Chiral Potts and a coupled Temperley--Lieb algebra}\label{sec:cp}
In this section we are showing that the spin-chain Hamiltonian of the three state
chiral Potts model can be re-expressed in terms of two coupled Temperly--Lieb algebras.
The three state chiral Potts spin chain Hamiltonian 
	(with periodic boundary conditions)
is \cite{Howes:1983mk}
\begin{equation}
H_{cp}=-\sum_{j=1}^L\sum_{n=1}^{2}\left(\alpha_n(X_j)^n+\bar{\alpha}_n(Z_j Z^{2}_{j+1})^n\right),
\end{equation}
where the operators $X_j$ and $Z_j$  satisfy
\begin{align}
	X_i^3 &=\mathbf{1} & Z_i^3 &=\mathbf{1} & X_iZ_i &=Z_iX_i\omega,\quad \omega:=e^{i2\pi/3}\\
	X_iX_j &= X_jX_i & Z_iZ_j &= Z_jZ_i & X_iZ_j &= Z_jX_i,\quad i\neq j.
\end{align}
	Periodic boundary conditions are imposed by interpreting the values of indices $i$, $j$ modulo $L$, i.e. we impose $X_{L+1}=X_1$ and $Z_{L+1}=Z_1$. Unless otherwise stated, we implicitly assume the spin chains appearing in the text having periodic boundary conditions, and will refer to such spin chains as \emph{closed}. By an \emph{open} chain is implied a spin chain with free boundary conditions. In the chiral Potts chain, the Hamiltonian of an open chain is obtained from $H_{cp}$ by dropping the terms $(Z_LZ^{2}_{L+1})^p$, $p=1,2$.

A commonly used convention is to define the parameters $\alpha$ and $\bar{\alpha}$ as
\begin{equation}
\alpha_n=\lambda\frac{e^{i(2n-3)\phi/3}}{\sin \pi n/3} \qquad \bar{\alpha}_n=\frac{e^{i(2n-3)\bar{\phi}/3}}{\sin \pi n/3} ,
\end{equation}
where, in the general case, $\phi$ and $\bar{\phi}$ are independent parameters.
Writing out the Hamiltonian more explicitly we then get
\begin{equation}
	H_{cp}=-\frac{2}{\sqrt{3}}\sum_{j=1}^L\left\{\lambda\left(e^{-i \phi/3}X_j+e^{i\phi/3}X_j^2\right)+ e^{-i\bar{\phi}/3}Z_jZ_{j+1}^2+e^{i\bar{\phi}/3}Z_j^2Z_{j+1}\right\}.
	\label{cpHam}
\end{equation}
The chiral Potts model is known \cite{AuYang:1987zc, 244547} to possess an R-matrix when its parameters are related as follows
\begin{equation}
\label{CP:int_manifold}
\lambda \cos \phi=\cos \bar{\phi}.
\end{equation}
We refer to the set of solutions to this equation in $\C^3$ as the integrability manifold for the chiral Potts model. Generic points on the integrability manifold correspond to spectral curves of higher genus.
For $\phi=\bar{\phi}=\pi/2$ (thus satisfying $\lambda\cos\phi=\cos\bar\phi$)  von Gehlen and Rittenberg \cite{vonGehlen:1984bi} showed that the Hamiltonian admits 
an infinite set of commuting conserved charges. Moreover, it follows from the results of \cite{Perk:1987} that it possesses a symmetry generated by the Onsager algebra, rending the model superintegrable.

Introduce the combinations $e_i$ and $f_i$ according to
\begin{align}
	e_{2i-1} &= 3^{-1/2}(\omega X_i+\omega^2 X_i^2 +1) & e_{2i} &= 3^{-1/2}(\omega Z_iZ_{i+1}^2+\omega^2Z_i^2Z_{i+1}+1)\label{edef}\\
	f_{2i-1} &= 3^{-1/2}(\omega^2 X_i+\omega X_i^2 +1) & f_{2i} &= 3^{-1/2}(\omega^2 Z_iZ_{i+1}^2+\omega Z_i^2Z_{i+1}+1).\label{fdef}
\end{align}
The following relations are straightforward to verify.
\begin{align}
	e_i^2 &= \sqrt{3}e_i & f_i^2 &= \sqrt{3} f_i & e_ie_j &= e_je_i & f_if_j &= f_j f_i & e_if_j &= f_j e_i\ \text{for } |i-j|>1\\
	e_ie_{i\pm1}e_i &= e_i & f_if_{i\pm 1}f_i &= f_i & e_if_{i\pm 1}e_i &= e_i & f_ie_{i\pm 1}f_i &= f_i & e_if_i &= 0 = f_ie_i
\end{align}
The operators $e_i$ and $f_i$ thus define two coupled Temperley--Lieb algebras.
Furthermore, the following four sets of quadratic relations (the first line expresses two sets) can be shown to follow from the definition of $e_i$ and $f_i$.
\begin{align}
&\pm \frac{i}{2}[e_i-f_i,e_{i\pm 1}+f_{i\pm 1}] + \frac{\sqrt{3}}{2}\{e_i + f_{i},e_{i\pm1}- f_{i\pm 1}\}  = 
2(e_{i\pm 1}-f_{i\pm 1}) \label{Nr1}\\
&\frac{i}{2}\{e_i-f_i,e_{i-1}-f_{i-1}\}-\frac{\sqrt{3}}{2}[e_{i-1}+f_{i-1},e_{i}+f_{i}]=0 \label{Nr2}\\
& \frac{3\sqrt{3}}{2}\{e_i+f_i,e_{i-1}+f_{i+1}\}-\frac{i}{2}[e_{i-1}-f_{i-1},e_{i}-f_i] 
-6  (f_{i-1}+e_{i-1}+f_i+e_i) +4 \sqrt{3}=0 \label{Nr3}
\end{align}
Here, $\{\cdot,\cdot\}$ denotes the anticommutator.
Using relations \eqref{Nr1} and \eqref{Nr2} (one can also use \eqref{Nr1} and \eqref{Nr3}, or \eqref{Nr2} and \eqref{Nr3})
one easily shows the following set of cubic relations.
\begin{align}
\label{mixedcubic1}
	f_ie_{i\pm 1}e_i &= \mp \frac{\omega-\omega^{-1}}{\sqrt{3}}\left(\omega^{\pm 1} e_{i\pm 1}e_i-f_{i\pm 1}e_i\right) +\omega^{\mp 1}e_i\\
	e_ie_{i\pm 1}f_i &= \pm\frac{\omega-\omega^{-1}}{\sqrt{3}}\left(\omega^{\mp 1}e_{i\pm 1}f_i - f_{i\pm 1}f_i\right) + \omega^{\pm 1}f_i\\
	f_if_{i\pm 1}e_i &= \mp\frac{\omega-\omega^{-1}}{\sqrt{3}}\left(e_{i\pm 1}e_i - \omega^{\mp 1}f_{i\pm 1}e_i\right) + \omega^{\pm 1}e_i\\
	e_if_{i\pm 1}f_i &= \pm\frac{\omega-\omega^{-1}}{\sqrt{3}}\left(e_{i\pm 1}f_i -\omega^{\pm 1}f_{i\pm 1}f_i\right)+\omega^{\mp 1}f_i.
\end{align}

\begin{align}
\label{mixedcubic2}
	f_ie_{i\pm 1}e_i &= \mp \frac{\omega-\omega^{-1}}{\sqrt{3}}\left(\omega^{\pm 1} f_ie_{i\pm 1}-f_if_{i\pm 1}\right) +\omega^{\mp 1}f_i\\
	e_ie_{i\pm 1}f_i &= \pm\frac{\omega-\omega^{-1}}{\sqrt{3}}\left(\omega^{\mp 1}e_ie_{i\pm 1} - e_if_{i\pm 1}\right) + \omega^{\pm 1}e_i\\
	f_if_{i\pm 1}e_i &= \mp\frac{\omega-\omega^{-1}}{\sqrt{3}}\left(f_ie_{i\pm 1} - \omega^{\mp 1}f_if_{i\pm 1}\right) + \omega^{\pm 1}f_i\\
	e_if_{i\pm 1}f_i &= \pm\frac{\omega-\omega^{-1}}{\sqrt{3}}\left(e_ie_{i\pm 1} -\omega^{\pm 1}e_if_{i\pm 1}\right)+\omega^{\mp 1}e_i.
\end{align}
In addition, relations \eqref{Nr1} and \eqref{Nr2} follow from the cubic relations, so these are in fact equivalent.

	For a closed chain, the indices $i$ and $j$ in definitions \eqref{edef} and \eqref{fdef} are interpreted modulo $L$. Consequently, in all the subsequent relations the indices $i$ and $j$ are interpreted modulo $2L$.  In the case of an open chain the elements $e_{2L}$ and $f_{2L}$ are absent, and the relations are the same as for a closed chain.

Writing out the Hamiltonian \eqref{cpHam} in the new generators we get 
\begin{equation}
	H_{cp} = -\frac{4}{3^{1/2}}\sum_{i=1}^{L}\left\{ \lambda\sin{\frac{\phi-2\pi}{3}} e_{2i-1} - \lambda\sin{\frac{\phi+2\pi}{3}} f_{2i-1} + \sin{\frac{\bar{\phi}-2\pi}{3}} e_{2i} - \sin{\frac{\bar{\phi}+2\pi}{3}} f_{2i}\right\}.
	\label{TLHam}
\end{equation}
Restricting to $\phi=\bar{\phi}$ and $\lambda=1$, where an affine quantum group governs the integrability \cite{Gomez}, the Hamiltonian takes the form
\begin{equation}
	H' = -\frac{4}{3^{1/2}}\sum_{i=1}^{2L}\left\{\sin{\frac{\phi-2\pi}{3}}e_i - \sin{\frac{\phi+2\pi}{3}} f_i\right\} = -\frac{4}{3^{1/2}}\sin{\frac{\phi-2\pi}{3}}\sum_{i=1}^{2L}\left\{e_i - K(\phi) f_i\right\},
	\label{sdHam}
\end{equation}
where \[K(\phi) = \frac{\sin{\frac{\phi+2\pi}{3}}}{\sin{\frac{\phi-2\pi}{3}}}. \]
In the superintegrable case, $\phi=\bar{\phi}=\pi/2$, we instead get
\begin{equation}
	H'' = -\frac{4}{3^{1/2}}\sum_{i=1}^L\left\{\lambda(e_{2i-1}+\frac{1}{2}f_{2i-1}) + (e_{2i}+\frac{1}{2}f_{2i}) \right\}.
\end{equation}
In the latter case it is convenient to redefine $X_i\mapsto \omega X_i$, leaving all properties of $X_i$ and $Z_i$ invariant, which instead 
results in the superintegrable Hamiltonian
\begin{equation}
	H''' = -\frac{4}{3^{1/2}}\sum_{i=1}^L\left\{ \lambda(e_{2i-1}-f_{2i-1}) + (e_{2i}-f_{2i})\right\}.
	\label{siHam}
\end{equation}
One remark concerning these Hamiltonians is in order. Considering the definitions \eqref{edef} and \eqref{fdef} it is not surprising that the odd and even operators appear asymmetrically. It is therefore quite remarkable that the self dual Hamiltonian with $\lambda=1$, \eqref{sdHam},  takes the form that it does
	with odd and even operators appearing completely symmetrically.

	For an open chain, the Hamiltonian \eqref{cpHam} again takes the forms \eqref{TLHam}--\eqref{siHam}, the only difference is that the terms involving the (non-existent) elements $e_{2L}$ and $f_{2L}$ are absent.

Let us now define a (slight) generalisation of the coupled Temperley--Lieb algebra of the three state chiral Potts chain. For $\gamma\in\C$, consider the unital associative algebra 
	generated by $e_i$ and $f_i$, $i=1,\ldots,n$, with relations
\begin{align}
	e_i^2 &= \gamma e_i & f_i^2 &= \gamma f_i & e_ie_j &= e_je_i & f_if_j &= f_j f_i & e_if_j &= f_j e_i,\ \text{for } |i-j|>1\label{rel1}\\
	e_ie_{i\pm1}e_i &= e_i & f_if_{i\pm 1}f_i &= f_i & e_if_{i\pm 1}e_i &= e_i & f_ie_{i\pm 1}f_i &= f_i & e_if_i &= 0 = f_ie_i.\label{rel2}
\end{align}
In these relations the indices are interpreted modulo $n$, and the algebra thus contains two TL algebras of closed type. There is an open analogue of this algebra obtained by dropping the generators $e_n$ and $f_n$, together with the relations involving these. The open algebra thus contains two TL algebras of conventional (open) type.
Both the closed and the open algebras so defined are infinite dimensional for $n>2$. To see this, consider the element $x=e_1e_2f_1f_2$. It is straightforward to check that $x^n$ cannot be reduced to a word of shorter length for any $n\in\Z_+$.

We would therefore like to consider a smaller algebra, and we do this by imposing additional relations. To this end, choose $\alpha\in\C\backslash\{0\}$ and define $\mu=\alpha-\alpha^{-1}$, $\gamma=\alpha+\alpha^{-1}$. Define the algebra $\tilde{\mathcal{A}}_n(\alpha)$
	(of closed type)
 by imposing the additional relations
\begin{equation}
	\mp\mu[e_i-f_i,e_{i\pm 1}+f_{i\pm 1}] +\gamma\{e_i+f_i,e_{i\pm 1}-f_{i\pm 1}\} +4(f_{i\pm 1}-e_{i\pm 1}) = 0\label{rel3}
\end{equation}
\begin{equation}
	\mu \{e_i-f_i,e_{i-1}-f_{i-1}\} - \gamma [e_{i-1}+f_{i-1},e_i+f_i] = 0.\label{rel4}
\end{equation}
As can be straighforwardly shown, the relation between $\mu$ and $\gamma$ is enforced, assuming $\gamma\neq 0$, lest the quadratic relations kill all the $e_i$ and $f_i$.
	Although we have no proof, we believe that the algebras $\tilde{A}_n(\alpha)$ are generically finite dimensional. For the, slightly degenerate, case of $n=2$ this is easily confirmed, and we have checked a few less trivial cases by computer.

It will be useful to consider an even further reduced version, $\mathcal{A}_n(\alpha)$, obtained by
 imposing the additional relations
\begin{equation}
	k_1\{e_i+f_i,e_{i-1}+f_{i-1}\}+k_2[e_{i-1}-f_{i-1},e_i-f_i]+k_3(e_i+f_i+e_{i-1}+f_{i-1})+k_4=0,
	\label{rel5}
\end{equation}
where
\begin{equation}
\begin{split}
	k_1 & =-3(\alpha+\alpha^{-1})+\alpha^3+\alpha^{-3}=\gamma(\gamma^2-6)\\
	k_2 & =\alpha -\alpha^{-1}-\alpha^3+\alpha^{-3}=-\mu(\mu^2+2)\\
	k_3 & =4(\alpha +\alpha^{-1})^2=4\gamma^2\\
	k_4 & =-8(\alpha+\alpha^{-1})(\alpha^2+\alpha^{-2})=-8\gamma (\gamma^2-2).
\end{split}
\end{equation}
For the value $\alpha=e^{-\pi i/6}$,
the relations \eqref{rel3} and \eqref{rel4} are equivalent to \eqref{Nr1} and \eqref{Nr2}
respectively, whereas the relations \eqref{rel5} are equivalent to \eqref{Nr3}. 
	It is straightforward to show that $\mathcal{A}_n(\alpha)$ is finite dimensional as long as $\alpha\neq \pm i$, and that it is trivial for $\alpha=\pm1$ 
	(note that for these values of $\alpha$, either $\mu$ or $\gamma$ vanishes, leading to a non-generic form of \eqref{rel3}, \eqref{rel4}, and \eqref{rel5}).

	There are of course also open versions of these algebras, $\tilde{\mathcal{A}}_n^{o}(\alpha)$ and $\mathcal{A}_n^o(\alpha)$ respectively, obtained by removing the generators $e_n$ and $f_n$ and the corresponding relations. The algebra $\tilde{\mathcal{A}}_n^o(\alpha)$, and therefore also $\mathcal{A}_n^o(\alpha)$, is finite dimensional for every $n\in\Z_+$ (this may, for instance, be shown by induction on $n$).

\section{A question of integrability}\label{sec:integrab}

We will now examine whether the algebra $\mathcal{A}_n(\alpha)$ leads to even more integrable models of the chiral Potts type.
	To this end we will consider Hamiltonians on closed chains which can be written in terms of some representation of an algebra $\mathcal{A}_n(\alpha)$. In particular, periodicity of the chain is enforced by the closed type of algebra together with the appearance of all generators in the Hamiltonian
For exactly solvable nearest neighbour spin chain models, the existence of an R-matrix guarantees the existence of an
infinite number of commuting charges (for chains of infinite length). It was shown by Tetelman \cite{Tetelman:1982} that when
the R-matrix satisfies a certain difference property with respect to the spectral parameters (corresponding to a spectral curve of genus one or less), 
these charges can be generated by a boost, or ladder, operator (see \cite{Sklyanin:1991ss} for a review).
	Consider a periodic chain with $L$ sites.
 If one starts out with a nearest neighbour Hamiltonian
\begin{equation}
H=\sum_{j=1}^{L} H_{j,j+1},
\end{equation}
then the boost operator $D$ is another local operator such that the quantities $Q_n$, $n\geq 0$, defined recursively by 
$$
Q_{n+1}=[D,Q_{n}] \quad \mbox{with} \quad Q_0:= H.
$$
form a set of mutually commuting charges. In fact, it is too restrictive to demand that $D$ is a well defined operator, it is enough that the derivation $[ D,\cdot\ ]$ is well defined.
For an XYZ spin chain one may define such a boost operator $D_0$ as 
\begin{equation}\label{initboost}
D_0=\sum_j jH_{j,j+1},
\end{equation}
and commutativity of the corresponding charges was shown in \cite{Fuchssteiner, Araki}.

Consider an abstract operator $H$ together with a derivation $\mathcal{D}$ of an operator algebra containing $H$. The operators $Q_n$ defined 
by $Q_0:= H$, $Q_{n+1}:=\mathcal{D}(Q_n)$, are mutually commuting if and only if $[Q_{n+1},Q_n]=0$ for all $n\geq 0$. This follows straightforwardly 
by repeated application of the Leibniz property of $\mathcal{D}$. In a given example one may hope to find an inductive 
proof of $[Q_{n+1},Q_n]=0$, $n\geq 0$, thus reducing the proof to showing $[Q_1,Q_0]=[\mathcal{D}(H),H]=0$. For the XYZ chain such a proof 
was outlined in \cite{Fuchssteiner}. An inductive step like that will necessarily depend on particular properties of a given example, or class
 of examples. However, we recall a conjecture (Conjecture 2 in \cite{Grabowski:1994rb}) claiming that in a periodic, translationally invariant quantum 
chain with nearest neighbour Hamiltonian $H$, the existence of an operator $D$ such that  $[D,H]$ is non--trivial and $[[D,H],H]=0$, is enough to ensure
 the commutativity of the charges $Q_n$ defined from the derivation $[D,\cdot\ ]$.

Adopting the sentiment behind the latter conjecture we will first search for a charge $Q$ commuting with a certain type of nearest neighbour
 Hamiltonian $H$  in terms of a representation of $\mathcal{A}_n(\alpha)$. As we will see, demanding the existence of such a charge gives a 
condition that for the chiral Potts case, $\mathcal{A}_n(e^{-\pi i/6})$, coincides with \eqref{CP:int_manifold}. We then show that there exists a 
derivation $[D,\cdot\ ]$, generalising $[D_0,\cdot\ ]$ from the XYZ chain, such that $Q\equiv Q_1=[D,H]$. Assuming the conjecture above holds, 
integrability then follows. Although we have not found a proof of $[Q_{n+1},Q_n]=0$ directly from the relations of $\mathcal{A}_n(\alpha)$, a computer
 calculation confirms that $[Q_2,Q_1]=0$.

Let us consider the Hamiltonian
\begin{equation}\label{genHam}
H=\sum_i \tilde{\lambda}_1[\delta_1 e_{2i-1}-\epsilon_1f_{2i-1}]+\tilde{\lambda}_2
[\delta_0 e_{2i}-\epsilon_0f_{2i}],
\end{equation}
where we have introduced
\begin{equation}
\delta_{1}=(\sin \phi/3 -k\cos \phi /3) \qquad \delta_{0}=(\sin \bar{\phi}/3 -k\cos \bar{\phi}/3)
\end{equation}
and
\begin{equation}
\epsilon_{1}=(\sin \phi/3 +k\cos \phi/3) \qquad \epsilon_{0}=(\sin \bar{\phi}/3 +k\cos \bar{\phi}/3),
\end{equation}
and where $e_i, f_i$ are assumed to belong to some representation of $\mathcal{A}_n(\alpha)$. 
In the sequel it will be convenient to leave the normalisation of both even and odd sites arbitrary, and we have done this by including the arbitrary
constants $\tilde{\lambda}_1$ and $\tilde{\lambda}_2$.
Note that for $k=\sqrt{3}$, $\tilde\lambda_1=2\lambda/\sqrt{3}$, $\tilde\lambda_2=2/\sqrt{3}$ this Hamiltonian takes the form \eqref{TLHam}.
Define 
$$Q_B:=[D_0, H],$$
where $D_0$ is defined as in \eqref{initboost}. In order to find a first commuting charge $Q$ we wish to find another operator $Q_E$ such that $Q=Q_B+Q_E$.
We make the nearest neighbour ansatz
$$
Q_E=\sum_i d_1(e_{2i-1}+f_{2i-1})+d_2(e_{2i-1}-f_{2i-1})+d_3(e_{2i}+f_{2i})+d_4(e_{2i}-f_{2i}),
$$
and try to determine the parameters $d_i$ by demanding
$$
[H,Q]\equiv [H,Q_B+Q_E]=0.
$$
Consider first  the simpler case when
$\phi=\bar\phi = 3\pi/2$
 (the case when the Hamiltonian takes the form \eqref{siHam}).
Then
\begin{equation}
\begin{split}
[H,Q_B]=- 3(\alpha-\alpha^{-1})\tilde{\lambda}_1 \tilde{\lambda}_2\sum_i(& \tilde{\lambda}_1[e_{2i}+f_{2i},e_{2i+1}-f_{2i+1}+e_{2i-1}-f_{2i-1}]+\\
&\tilde{\lambda}_2[e_{2i+1}+f_{2i+1}+e_{2i+1}+f_{2i+1},e_{2i}-f_{2i}]).
\end{split}
\end{equation}
From this we conclude that
\begin{equation}
Q_E=-3  (\alpha-\alpha^{-1})\tilde{\lambda}_1\tilde{\lambda}_2\sum_i (e_{2i}+f_{2i}+e_{2i+1}+f_{2i+1}).
\end{equation}
In the general case we have
\begin{equation}
\begin{split}
&[H,Q_B]= \sum_i a_1 [e_{2i}-f_{2i},e_{2i+1}-f_{2i+1}+e_{2i-1}-f_{2i-1}]+\\
& a_2 [e_{2i}+f_{2i},e_{2i+1}+f_{2i+1}+e_{2i-1}+f_{2i-1}]\\
& +a_3 [e_{2i}-f_{2i},e_{2i+1}+f_{2i+1}+e_{2i-1}+f_{2i-1}]\\
&+a_4 [e_{2i}+f_{2i},e_{2i+1}-f_{2i+1}+e_{2i-1}-f_{2i-1}],
\end{split}
\end{equation}
with
\begin{equation}
\begin{split}
&a_1= \tilde{\lambda}_1\tilde{\lambda}_2k \left(\tilde{\lambda}_1(k^2\cos^2 {\phi/3}-3\sin^2 {\phi/3})\cos\bar{\phi}/3-
\tilde{\lambda}_2(k^2\cos^2 {\bar{\phi}/3}-3\sin^2 {\bar{\phi}/3})\cos{\phi/3}\right)
\frac{(\alpha-\alpha^{-1})(\alpha^2+\alpha^{-2})}{\alpha^2+\alpha^{-2}-4} \\
& a_2=2k \tilde{\lambda}_1\tilde{\lambda}_2(- \tilde{\lambda}_1\cos\phi/3+\tilde{\lambda}_2\cos \bar{\phi}/3)\sin \phi/3 \sin \bar{\phi}/3
\frac{(\alpha+\alpha^{-1})^2}{\alpha-\alpha^{-1}}\\
&a_3= \tilde{\lambda}_1\tilde{\lambda}_2(\tilde{\lambda}_2(k^2\cos^2 {\bar{\phi}/3}-3\sin^2 {\bar{\phi}/3}) 
+2k^2\tilde{\lambda}_1\cos \bar{\phi}/3 \cos {\phi/3})\sin {\phi/3}(\alpha-\alpha^{-1})\\
&a_4=- \tilde{\lambda}_1\tilde{\lambda}_2(\tilde{\lambda}_1(k^2\cos^2 {\phi/3}-3\sin^2 {\phi/3}) +2k^2\tilde{\lambda}_2\cos \phi/3 \cos \bar{\phi}/3)\sin \bar{\phi}/3(\alpha-\alpha^{-1}).
\end{split}
\end{equation}
The solution for $Q_E$ is then 
\begin{equation}\label{sol}
\begin{split}
& d_1=2\tilde{\lambda}_1\tilde{\lambda}_2\sin {\phi/3}\sin \bar{\phi}/3\frac{(\alpha+\alpha^{-1})^2}{\alpha-\alpha^{-1}}
\\
& d_2=-\tilde{\lambda}_1\frac1{k}\frac{\sin(\bar{\phi}/3)}{\cos(\bar{\phi}/3)} (\alpha-\alpha^{-1})
\left(\tilde{\lambda}_1((k^2+3-2\left(\frac{\alpha+\alpha^{-1}}{\alpha-\alpha^{-1}}\right)^2)\cos^2({\phi/3})
+ 2 \left(\frac{\alpha+\alpha^{-1}}{\alpha-\alpha^{-1}}\right)^2-3)+2 \tilde{\lambda}_2 k^2\cos(\phi/3)\cos(\bar{\phi}/3)\right)\\
&  d_3=2 \tilde{\lambda}_1\tilde{\lambda}_2\sin \bar{\phi}/3\sin {\phi/3}\frac{(\alpha+\alpha^{-1})^2}{\alpha-\alpha^{-1}}
 \\
& d_4=-\tilde{\lambda}_2\frac1{k}\frac{\sin(\phi/3)}{\cos(\phi/3)} (\alpha-\alpha^{-1})
\left(\tilde{\lambda}_2((k^2+3-2\left(\frac{\alpha+\alpha^{-1}}{\alpha-\alpha^{-1}}\right)^2)\cos^2(\bar{\phi}/3)+2\left(\frac{\alpha+\alpha^{-1}}{\alpha-\alpha^{-1}}\right)^2-3+2 \tilde{\lambda}_1k^2\cos(\phi/3)\cos(\bar{\phi}/3)\right)\\
\end{split}
\end{equation}
together with the relation 
\begin{equation}
\label{cond}
a_1=d_4 \tilde{\lambda}_1\sin \phi/3-d_2\tilde{\lambda}_2\sin \bar{\phi}/3.
\end{equation}
The latter relation can be written as an equation of the form
\begin{equation}\label{intcondition}
\tilde{\lambda}_1 f(\phi, \bar{\phi})=\tilde{\lambda}_2 f(\bar{\phi},\phi),
\end{equation}
with
\begin{equation}\label{functionf}
\begin{split}
&f(\bar{\phi},{\phi})=\frac1{3}\left((C_{0} k^{4} + 3 \, C_{0} k^{2} - 2 \, k^{2}+C_1) \cos\left(\bar{\phi}/3\right)^{2} \cos\left(\phi/3\right)^{2} \right. \\
&\left. +(2k^2- 3 \, C_{0} k^{2}+ 2\,C_2-3) \cos\left(\phi/3\right)^{2}  - C_{1} \cos\left(\bar{\phi}/3\right)^{2} - (2 C_{2}-3)\right) \cos\left(\bar{\phi}/3\right),
\end{split}
\end{equation}
(the factor $1/3$ has been chosen such that $f(\bar{\phi},{\phi})$ reduces to $\cos \bar{\phi}$ for $k=\sqrt{3}$ and $\alpha=e^{-\pi i/6}$) where
\be
\label{Cparameters}
C_0=\frac{\alpha^{2} + \alpha^{-2}}{\alpha^{2} + \alpha^{-2} - 4} \qquad C_1=k^{2} - \frac{2 \, {\left(\alpha + \alpha^{-1}\right)}^{2}}{{\left(\alpha - \alpha^{-1}\right)}^{2}} + 3
\qquad C_2=\frac{ {\left(\alpha + \alpha^{-1}\right)}^{2}}{{\left(\alpha - \alpha^{-1}\right)}^{2}}.
\ee
It is straightforward to check that for $k=\sqrt{3}$ and $\alpha=e^{-\pi i/6}$ (i.e. the chiral Potts case) this reduces to
\begin{equation}
\lambda\cos\phi=\cos\bar{\phi} \qquad \lambda=\frac{\tilde{\lambda}_1}{\tilde{\lambda}_2}.
\end{equation}
Thus, in the chiral Potts case the condition of having a ``first'' commuting charge coincides with the condition for the model to possess an
R-matrix. We will therefore interpret \eqref{cond} as an integrability condition for the Hamiltonian \eqref{genHam}. Note, however, that the 
solution \eqref{sol} does not exist when $\alpha=\pm 1$, $\phi=n3\pi/2$, or $\bar\phi=n3\pi/2$.

Next we will show that there exists an operator $D$, a generalisation of the boost operator, such that we can write 
	$Q\equiv Q_1=[D,H]$, where $H$ is the Hamiltonian in \eqref{genHam} with the choice $\tilde{\lambda}_1=f(\bar{\phi},\phi)$ and $\tilde{\lambda}_2=f(\phi,\bar{\phi})$.  Let us write the Hamiltonian as
\begin{equation}
H =\sum_i h_{i,i+1},
\end{equation}
\begin{equation}\label{kparam}
\begin{split}
&h_{2i-1,2i}(k)= -kf(\bar{\phi},\phi)\cos {\phi/3} (e_{2i-1}+f_{2i-1})+f(\bar{\phi},\phi)\sin{\phi/3} (e_{2i-1}- f_{2i-1})\\
&h_{2i,2i+1}(k)= -k f(\phi,\bar{\phi}) \cos {\bar{\phi}/3}(e_{2i}+f_{2i})+f(\phi,\bar{\phi})\sin{\bar{\phi}/3}(e_{2i}-f_{2i}),
\end{split}
\end{equation}
where the function $f$ is written out in \eqref{functionf}. As we have seen, the existence of a first commuting charge implies the condition \eqref{intcondition} which now is identically satisfied because of our choice of $\tilde{\lambda}_1$ and $\tilde{\lambda}_2$. 
We have already shown that the first commuting charge is of the form
\begin{equation}
Q_1=[D_0,H]+Q_E,
\end{equation}
for a nearest neighbour term  $Q_E$.
We want to show that $Q_E$ can be obtained by taking derivatives of  $H$ with respect to the parameters according to
\begin{equation}
\label{charge_cond}
Q_E= D_E H, \qquad \mbox{with} \qquad D_E:=\beta\partial_{\phi}+\bar{\beta}\partial_{\bar{\phi}}.
\end{equation}
	Interpreting the condition \eqref{charge_cond} as an equation for $\beta$, $\bar\beta$, and $k$, it turns out that this equation can be solved (see appendix \ref{app:boost} for details). We will not write out the general solution for $\beta$ and $\bar\beta$ (these can be found from equations \eqref{beta} and \eqref{barbeta}). The equation for $k$ reads
\begin{equation}\label{keq}
\begin{split}
& f(\phi,\bar{\phi})f(\bar{\phi},\phi)(d_{3ex}\cos \bar{\phi}/3-d_{4kex}\sin \bar{\phi}/3) -
f(\phi,\bar{\phi})f(\bar{\phi},\phi)(d_{1ex}\cos {\phi/3}-d_{2kex}\sin {\phi/3})-\\
&(d_{3ex}\sin \bar{\phi}/3+ d_{4kex}\cos \bar{\phi}/3 )\left(f(\phi,\bar{\phi})\partial_{\bar{\phi}}f(\bar{\phi},\phi)-
f(\bar{\phi},\phi)\partial_{\bar{\phi}} f(\phi, \bar{\phi})\right)+\\
& (d_{1ex}\sin {\phi/3}+ d_{2kex}\cos {\phi/3} )\left(f(\bar{\phi},\phi)\partial_{\bar{\phi}}f(\phi,\bar{\phi})-
f(\phi, \bar{\phi} )\partial_{\bar{\phi}} f(\bar{\phi},\phi)\right)=0,
\end{split}
\end{equation}
where the $d_{1ex}$ are defined in equation \eqref{eq:dex}. Note that the equation above does not have any $k$'s in the denominator. The degree of $f(\phi,\bar{\phi})$ in terms of $k^2$ is two, while $d_{1ex}$, $d_{3ex}$,
$d_{2kex}$ and $d_{4kex}$ are all of degree two (it is not obvious that $d_{2kex}$ and $d_{4kex}$ are of this order
but it has been checked that the degree four terms cancel out). 
Thus one would expect that this equation is of order six in $k^2$.
However, the terms of order six, and zero both vanish, as has been checked using sage math. 
\begin{figure*}[t]
\centering
\includegraphics[width=0.5\textwidth]{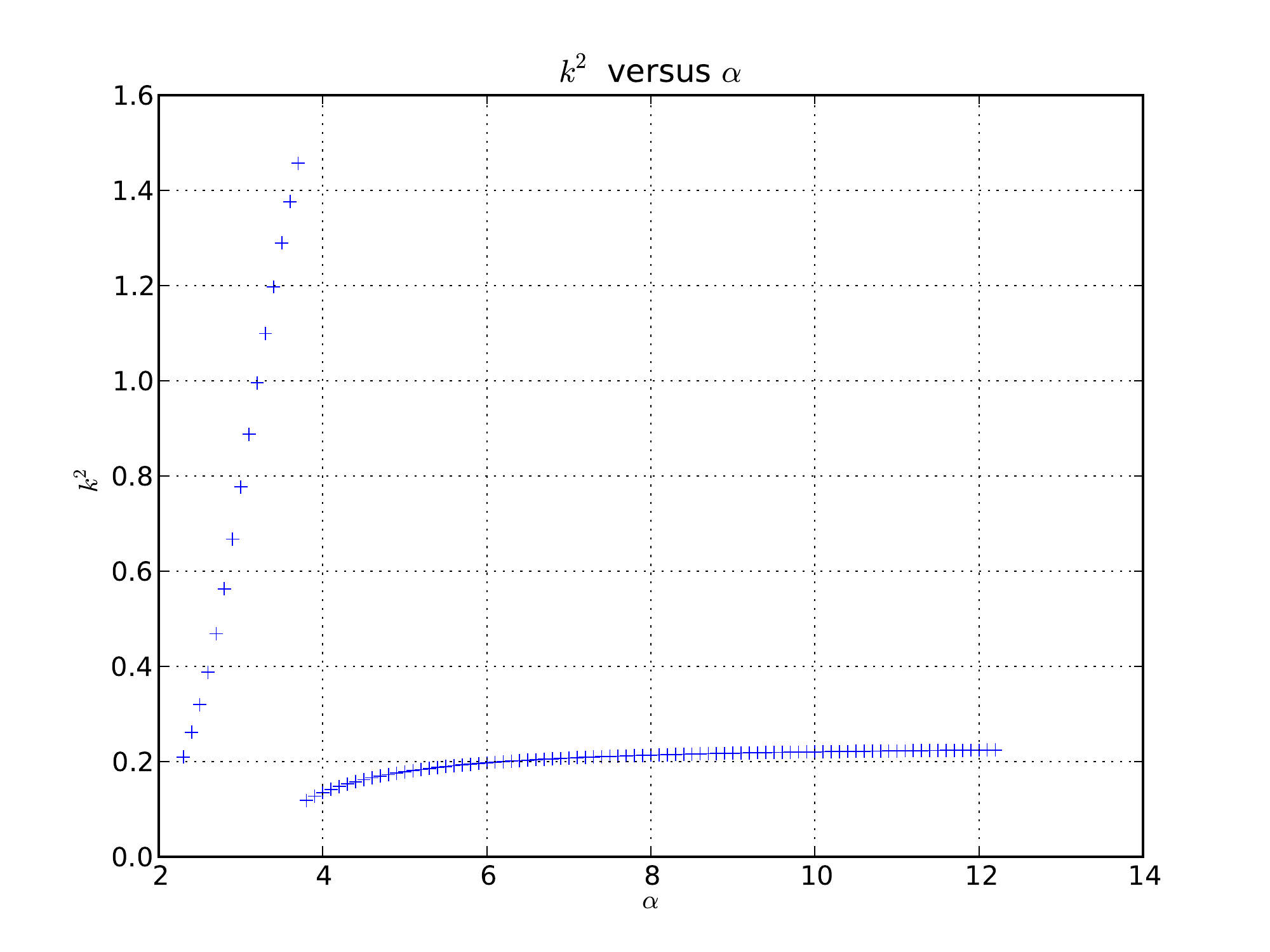}
	\caption{The $\alpha$-dependence of one of the four roots $k^2$ of eq. (52) for $\phi=\pi/7$ and $\bar\phi=\pi / 3$.}
\label{figur}
\end{figure*}
This means that we have an equation of degree four for $k^2$. A solution to the equation for the particular case when $\phi=\pi/7$ and $\bar{\phi}=\pi/3$ can be seen in figure (\ref{figur}) as a function of $\alpha$.
To summarize, we have now shown that we can generate the first charge of the Hamiltonian \eqref{kparam}, with the parameter $k$ solving the equation \eqref{keq}, as follows.
\begin{equation}
Q_1=[D, H]   \qquad \mbox{and}\qquad D(k)=\sum_i i H_{i,i+1}+D_E
\end{equation}
Here, $D_E$ is defined as in \eqref{charge_cond} where $\beta$ and $\bar\beta$ are obtained from equations \eqref{beta} and \eqref{barbeta}.

Interestingly, in the case $\alpha=e^{-\pi i/6}$ of the chiral Potts model, the equation \eqref{keq} is identically satisfied for all $k$. 
In this case $k$ is thus a free parameter, and for the choice $k=\sqrt{3}$ (corresponding to the chiral Potts Hamiltonian) the form of $D_E$ simplifies
\begin{equation}
D_E=\beta\partial_{\phi}+\bar{\beta}\partial_{\bar{\phi}}
\qquad \beta=i3\sqrt{3}\cos{\phi}\sin{\bar{\phi}} \qquad \bar{\beta}=i3\sqrt{3}\cos{\bar{\phi}}\sin{{\phi}}.
\end{equation}

By the conjecture of \cite{Grabowski:1994rb} we expect the charges $Q_n$, defined recursively using $D$, to form (for an infinite chain) an
 infinite set of commuting conserved charges. So far we have not been able to provide a general proof of this. However, we have checked by computer
 that $[Q_2,Q_1]=0$. Using the derivation property of $[D(k),\cdot]$ it furthermore follows straightforwardly 
that $[Q_4,H]=0=[Q_3,Q_1]=[Q_3,H]=[Q_2,H]$.

We may now use our generalised boost operator to get explicit forms of the charges.
 For instance, as we have already seen,
\begin{equation}
Q_1=-\sum_i [h_{i,i+1},h_{i+1,i+2}]+D_E H,
\end{equation}
and furthermore
\begin{equation}
\begin{split}
&Q_2=\sum_i 2[h_{i,i+1},[h_{i+1,i+2},h_{i+2,i+3}]]+[h_{i,i+1}[h_{i,i+1},h_{i+1,i+2}]]-X_i- \\
& 2[h_{i,i+1},D_Eh_{i+1,i+2}]+[h_{i+1,i+2},D_Eh_{i,i+1}] +D_ED_EH,
\end{split}
\end{equation}
where $X_i$ is defined from 
\begin{equation}
[h_{i,i+1},[h_{i,i+1},h_{i+1,i+2}]-D_E h_{i+1,i+2}]+[h_{i+1,i+2},[h_{i,i+1},h_{i+1,i+2}]-D_E h_{i,i+1}]=X_i-X_{i+1}.
\end{equation}
This last equation is a generalisation of the Reshetikhin criterium.

In the Onsager integrable case, $\lambda$ is a free parameter, and using the original form of the Hamiltonian ($\tilde{\lambda}_1=\lambda$ and
$\tilde{\lambda}_2=1$), the chiral Potts boost operator can be written as
\begin{equation}
 D_E=-i3\sqrt{3}(\partial_{\phi}+\lambda \partial_{\bar{\phi}}).
\end{equation}
This implies
\begin{equation}
 D_ED_E H=\sum_i3\lambda ((e_{2i-1}-f_{2i-1})+\lambda (e_{2i}-f_{2i})
\end{equation}
and
\begin{equation}
 X_{2i}=-6\lambda ((e_{2i-1}-f_{2i-1})+\lambda (e_{2i}-f_{2i}),
\end{equation}
leading to
\begin{equation}
\begin{split}
&Q_2=\sum_i 2[h_{i,i+1},[h_{i+1,i+2},h_{i+2,i+3}]]+[h_{i,i+1}[h_{i,i+1},h_{i+1,i+2}]]-2[h_{i,i+1},D_Eh_{i+1,i+2}]+[h_{i+1,i+2},D_Eh_{i,i+1}]+ \\
& 9\lambda ((e_{2i-1}-f_{2i-1})+\lambda (e_{2i}-f_{2i}).
\end{split}
\end{equation}

\section{The Onsager algebra and \texorpdfstring{$\mathcal{A}_n(\alpha)$}{An(a)}}\label{sec:Onsager}
In Onsagers first solution of the Ising model an algebra, now known as the Onsager algebra, played a central
role. The Onsager algebra is spanned by generators $A_n$, $G_n$, $n\in\Z$, satisfying
$$
[A_m,A_l]=4 G_{m-l} \qquad [A_m, G_l]=2(A_{m-1}-A_{m+l}).
\qquad [G_m,G_l]=0
$$
Consider a Hamiltonian of the form
\begin{equation}\label{supintHam}
H\propto A_0+\lambda A_1
\end{equation}
If $A_0$ and $A_1$ are generators of the Onsager algebra, one can construct
charges commuting with the Hamiltonian. In order for a model with Hamiltonian \eqref{supintHam} to exhibit an Onsager algebra, it is enough that $A_0$ and $A_1$ satisfy
the Dolan-Grady condition:
\begin{equation}
\label{Dolan_Grady}
[A_1,[A_1,[A_1,A_0]]]=16 [A_1,A_0] \qquad [A_0,[A_0,[A_0,A_1]]]=16 [A_0,A_1].
\end{equation}
The factor of $16$ is due to a particular normalization.
Since the chiral Potts model satisfies this at the special point $\phi=\bar{\phi}=\pi/2$,
it is natural to ask is if the algebra $\mathcal{A}_n(\alpha)$ can be related to the Onsager algebra in greater generality. 

Let us make an ansatz for the $A_0$ and $A_1$
\begin{equation}
A_0=k\sum_i (e_{2i}-f_{2i}) \qquad A_1=k\sum_i (e_{2i+1}-f_{2i+1}),
\end{equation}
where $e_i$ and $f_i$ lie in a representation of $\mathcal{A}_n(\alpha)$. Of course, any Hamiltonian of the form \eqref{supintHam} is Onsager integrable if there is a choice of $k$ such that $A_0$ and $A_1$ satisfy \eqref{Dolan_Grady}.
Adopting periodic boundary conditions, and disregarding the cases $\gamma=\pm2$ or $\gamma=0$ (which need separate
treatment) and $\gamma=\pm\sqrt{6}$ (where our analysis does not apply), we get
\begin{equation}
\begin{split}\label{genericDG}
&[A_1,[A_1,[A_1,A_0]]]=k^{-2}(12+36\frac{(\alpha-\alpha^{-1})^2}{\alpha^2+\alpha^{-2}-4}+\gamma^2 )[A_1,A_0] \\
&[A_0,[A_0,[A_0,A_1]]]=k^{-2}(12+36\frac{(\alpha-\alpha^{-1})^2}{\alpha^2+\alpha^{-2}-4}+\gamma^2 )[A_0,A_1] .
\end{split}
\end{equation}
See appendix \ref{app:onsager} for technical details. From this we conclude that the Dolan--Grady condition is satisfied when
$$
k=\frac{1}{4}\left( 12+36\frac{(\alpha-\alpha^{-1})^2}{\alpha^2+\alpha^{-2}-4}+\gamma^2  \right)^{1/2},
$$
	as long as the right hand side of this expression does not vanish. The vanishing of the RHS is a fourth order equation in $\alpha^2$ whose solutions do not coincide with any of the cases we already exempted from treatment, and thus there are an additional eight values of $\alpha$ where $\mathcal{A}_n(\alpha)$ does not imply the Dolan--Grady condition:
	\begin{equation}
		\alpha^2= -\frac{23\pm 3\sqrt{73}}{2} \pm\sqrt{\frac{3}{2}(197-23\sqrt{73})},
	\end{equation}
	where the two choices of sign are independent.
In the generic case, however, one can construct a representation of the Onsager algebra from any representation of the algebra $\mathcal{A}_n(\alpha)$. 
Notice that the case
$\mathcal{A}_n(e^{-\pi i/6})$ implies
\begin{equation}
k=\frac{3\sqrt{3}}{4}.
\end{equation}

Let us now consider the special case of $\alpha=\pm 1$ ($\Rightarrow \gamma=\pm 2$, $\mu=0$).
	 In this case we neither need to assert periodicity, nor do we need to impose the relations \eqref{rel5} in order to check the Dolan-Grady condition. In other words, we may work with either of the (closed respectively open) algebras $\tilde{A}_n(\alpha)$ or $\tilde{A}_n^o(\alpha)$.
The calculations simplify and we get
\begin{equation}
[A_1,[A_1,[A_1,A_0]]]=16 k^2[A_1,A_0] \qquad
[A_0,[A_0,[A_0,A_1]]]=16 k^2[A_0,A_1],
\end{equation}
implying $k=1$.
The other special case, $\alpha=\pm i$, also does not require periodicity or \eqref{rel5}.
The calculations again simplify, and we get
\begin{equation}
[A_1,[A_1,[A_1,A_0]]]=24 k^{-2}[A_1,A_0] \qquad
[A_0,[A_0,[A_0,A_1]]]=24 k^{-2}[A_0,A_1],
\end{equation}
implying $k=\sqrt{3/2}$.
For details of the calculations we refer to appendix \ref{app:onsager}.

One remark is in order. The factors of $\alpha^2+\alpha^{-2}-4$ in the denominators of \eqref{genericDG} of course excludes $\alpha^2=2\pm\sqrt{3}$, i.e. $\gamma=\pm\sqrt{6}$. In fact the whole analysis fails in this case, not only the final formula. This can be traced back to vanishing of the constant $k_1$ in \eqref{rel5}. We have seen no other signs of the corresponding values of $\alpha$ leading to special properties of $\mathcal{A}_n(\alpha)$ or of corresponding physical models.

\section{Some realisations of coupled Temperley-Lieb algebras}\label{sec:realisations}
\subsection{The staggered XXZ model as a representation of \texorpdfstring{$\tilde{\mathcal{A}}_n(1)$}{A~n(1)}}
In this section we will see that in fact the staggered/alternating isotropic XY Heisenberg spin chain is
superintegrable and that the staggered/alternating XXZ Heisenberg spin chain can be written with generators
of $\tilde{\mathcal{A}}_n(1)$. First let us write down the conventional spin $\frac{1}{2}$ representation of the Temperley-Lieb algebra with $e_i^2=2e_i$
 expressed in terms of Pauli matrices
\begin{equation}\label{XXZrepe}
e_i=(\mathbf{1}-\sigma_i^z\sigma_{i+1}^z+\sigma_i^x\sigma_{i+1}^x+\sigma_i^y\sigma_{i+1}^y)/2,
\end{equation}
\begin{equation}
\sigma^x=\left(\begin{array}{cc}  0 & 1 \\ 1 & 0 
\end{array} \right) \qquad
\sigma^y=\left(\begin{array}{cc}  0 & -i \\ i & 0 
\end{array} \right) \qquad
\sigma^z=\left(\begin{array}{cc}  1 & 0 \\ 0 & -1 
\end{array} \right) \,.
\end{equation}
Define $f_i$ as
\begin{equation}\label{XXZrepf}
f_i=(\mathbf{1}-\sigma_i^z\sigma_{i+1}^z-\sigma_i^x\sigma_{i+1}^x-\sigma_i^y\sigma_{i+1}^y)/2 \,.
\end{equation}
It is straightforward to verify that $f_i$ and $e_i$ satisfy the relations \eqref{rel1}, \eqref{rel2}, \eqref{rel3} and \eqref{rel4}, with $\alpha=1$.
 Notice that \eqref{rel5} is not satisfied, but as we have seen we nevertheless get a representation of the Onsager algebra from $\tilde{\mathcal{A}}_n(1)$.
With these definitions we have
\begin{equation}
e_i+kf_i= ((1+k)(\mathbf{1}-\sigma_i^z\sigma_{i+1}^z)+(1-k)\left(\sigma_i^x\sigma_{i+1}^x+\sigma_i^y\sigma_{i+1}^y\right))/2.
\end{equation}
This interaction term is exactly the interaction term of the XXZ Heisenberg spin chain.
We consider a
	periodic
 four parameter Hamiltonian
$$
H=\sum_i \lambda_1 (e_{2i}+k_1 f_{2i})+ \lambda_2 (e_{2i+1}+k_2 f_{2i+1}),
$$
which 
	in the XXZ representation \eqref{XXZrepe}, \eqref{XXZrepf}
 takes the following form.
\begin{equation}
\begin{split}
&H=\sum_i \lambda_1 (e_{2i}+k_1 f_{2i})+ \lambda_2 (e_{2i+1}+k_2 f_{2i+1})=\\ &\sum_i 
\lambda_1 ((1+k_1)(\mathbf{1}-\sigma_{2i}^z\sigma_{2i+1}^z)+(1-k_1)\left(\sigma_{2i}^x\sigma_{2i+1}^x+\sigma_{2i}^y\sigma_{2i+1}^y\right))/2\\
&+\lambda_2 ((1+k_2)(\mathbf{1}-\sigma_{2i+1}^z\sigma_{2i+2}^z)+(1-k_2)\left(\sigma_{2i+1}^x\sigma_{2i+2}^x+\sigma_{2i+1}^y\sigma_{2i+2}^y\right))/2
\end{split}
\end{equation}
The superintegrable case corresponds to $k_1=k_2=-1$, and the model then reduces to the staggered XX Heisenberg model (isotropic XY).
A direct proof that this model is integrable was presented in \cite{Capel:1975}, but to the best of our knowledge it was not previously known to be superintegrable.
Another special case of the staggered XXZ model above has previously been shown to be integrable:
\begin{equation}
H=\sum_i 
((1+k_1)(\mathbf{1}+\sigma_{i}^z\sigma_{i+1}^z)+(1-k_1)(-1)^i\left(\sigma_{i}^x\sigma_{i+1}^x+\sigma_{i}^y\sigma_{i+1}^y\right))/2
\end{equation}
This was shown in \cite{Capel:1976} to be mapped to a particular case of the XXZ model with Dzyaloshinski-Moriya 
interaction \cite{Dzyaloshinski:1958,Moriya:1969, Kontorovich:1967}
\begin{equation}
H=\sum_i \sigma_{i}^x\sigma_{i+1}^y-\sigma_{i}^y\sigma_{i+1}^x+\lambda_z \sigma_{i}^z\sigma_{i+1}^z,
\end{equation}
which is known to be integrable \cite{Alcaraz:1990}.

\subsection{A coupled Temperley-Lieb algebra for \texorpdfstring{$SU(N)$}{SU(N)} spin chains}
We can easily find a large class of models related to the algebra described by the relations \eqref{rel1} and \eqref{rel2}. Denote by $\{J_\alpha^{\phantom{\alpha}\beta}\}_{\alpha,\beta=1}^N$ a basis of $\mathfrak{gl}_N$ in the representation $N$ restricting to the defining representation of $\mathfrak{sl}_N$, and where $T:=\sum_\alpha J_\alpha^{\phantom{\alpha}\alpha}$ acts as the identity, with the commutators
$$[J_\alpha^{\phantom{\alpha}\beta},J_\gamma^{\phantom{\gamma}\delta}]=\delta_{\gamma}^\beta J_\alpha^{\phantom{\alpha}\delta}-\delta_\alpha^\delta J_\gamma^{\phantom{\gamma}\beta}.$$
Furthermore, let $\bar{J}_\alpha^{\phantom{\alpha}\beta}$ denote the same basis in the contragredient representation $N^+$.
It is straightforward to check that these representations satisfy
\bea
	J_\alpha^{\phantom{\alpha}\beta}J_\gamma^{\phantom{\gamma}\delta} & = & \delta_\gamma^\beta J_\alpha^{\phantom{\alpha}\delta}\label{Nrel}\\
	\bar{J}_\alpha^{\phantom{\alpha}\beta}\bar{J}_\gamma^{\phantom{\gamma}\delta} & = & -\delta_\alpha^\delta\bar{J}_\alpha^{\phantom{\alpha}\delta}.\label{NCrel}
\eea
Consider a chain where odd sites contain the representation $N$ and even sites contain its contragredient $N^+$. Define for each site $i$ the operator $e_i$ as
\begin{equation}
	e_i :=
\begin{cases}
	- \sum_{\alpha,\beta}J_{i,\alpha}^{\phantom{i,\alpha}\beta}\otimes \bar{J}_{i+1,\beta}^{\phantom{i+1,\beta}\alpha} & \text{if } i \text{ is odd}\\
	- \sum_{\alpha,\beta}\bar{J}_{i,\alpha}^{\phantom{i,\alpha}\beta}\otimes J_{i+1,\beta}^{\phantom{i+1,\beta}\alpha} & \text{if } i \text{ is even}.
\end{cases}
\end{equation}
These then satisfy the defining relations of a Temperley-Lieb algebra. Note that the $e_i$ are $SU(N)$ (or rather $GL(N)$) invariant. Such chains with nearest neighbour Hamiltonians expressed in terms of the $e_i$, i.e. the most general $SU(N)$ invariant nearest neighbour Hamiltonians
	on chains with alternating representations $N$ and $N^+$,
 were studied in \cite{Affleck} using an oscillator realisation. See also \cite{ReadSaleur} where the symmetries of these models were studied.
Using the notation $\xi=e^{2\pi i/N}$ we define another set of operators as
\begin{equation}
	f_i :=
\begin{cases}
	- \sum_{\alpha,\beta}\xi^{\alpha-\beta}J_{i,\alpha}^{\phantom{i,\alpha}\beta}\otimes \bar{J}_{i+1,\beta}^{\phantom{i+1,\beta}\alpha} & \text{if } i \text{ is odd}\\
	- \sum_{\alpha,\beta}\xi^{\alpha-\beta}\bar{J}_{i,\alpha}^{\phantom{i,\alpha}\beta}\otimes J_{i+1,\beta}^{\phantom{i+1,\beta}\alpha} & \text{if } i \text{ is even}.
\end{cases}
\end{equation}
Unlike $e_i$, the operators $f_i$ only commute with a Cartan subalgebra spanned by elements $J_\alpha^{\phantom{\alpha}\alpha}$ (no summation).
A straightforward calculation confirms that $e_i$ and $f_i$ satisfy \eqref{rel1} and \eqref{rel2} with $\gamma=N$. They do not, however, satisfy (all of the) relations \eqref{rel3}, \eqref{rel4}, \eqref{rel5}. Note, however, that with the conventional inner product on the representations, such that $(J_\alpha^{\phantom{\alpha}\beta})^\dagger=J_\beta^{\phantom{\beta}\alpha}$, both $e_i$ and $f_i$ are Hermitean.
	A $SU(N)$--invariant nearest neighbour Hamiltonian of the form
$$
	H_0 = \sum_i \lambda_i e_i,
$$
for some choice of couplings $\lambda_i$, can now be generalised to a $U(1)^{N-1}$--invariant Hamiltonian

\begin{equation}\label{genSUNHam}
	H=H_0+\sum_i \kappa_i f_i.
\end{equation}

\section{Discussion}\label{sec:disc}

The results in this paper indicate that a study of representations of $\mathcal{A}_n(\alpha)$ and $\tilde{\mathcal{A}}_n(\alpha)$ may be fruitful. Using any representation one can immediately write down integrable, and even superintegrable, Hamiltonians. Due to the algebraic structure we expect such models to have similar physical behaviour to the chiral Potts model.
We are, however, left with several open questions.

So far we did not investigate in detail the structure of the algebras $\mathcal{A}_n(\alpha)$ and $\tilde{\mathcal{A}}_n(\alpha)$. It is not difficult
 to see that the dimension of $\mathcal{A}_2(\alpha)=\mathcal{A}_3^o(\alpha)$ (for generic $\alpha$) is $9$. We have not determined the dimension for general $n$, however, and this appears not to be completely straightforward. We also do not know for which values of $\alpha$ the algebras are simple, semisimple respectively non-semisimple.

A representation of a conventional Temperley-Lieb algebra automatically gives an R-matrix. It would be very interesting to find a way to construct
 R-matrices from representations of $\mathcal{A}_n(\alpha)$, and to compare with the R-matrix of the chiral Potts model.

We have in this paper restricted ourselves to the three state model, but there exist natural generalisations to an arbitrary $N$ state chiral Potts chain resulting in $N-1$ types of TL generators.

One important open question is how to prove that the charges produced by our conjectured generalisation of a boost operator are conserved and mutually
 commuting. In known examples, the boost operator is directly related to corner transfer matrices (CTM's) \cite{Jimbo}, and Baxters method to
 calculate order parameters using the CTM \cite{Baxter:1977,Baxter:1981} has proved useful in both the Ising model and the XYZ model. There have been 
several attempts to apply Baxter's CTM method to the chiral Potts model, but it has been explained \cite{Baxter:2007} how the lack of difference property 
makes it impossible to use the same technique. If it can indeed be shown that our candidate is a suitable generalisation of a boost operator, one may speculate that this could analogously be related to a suitable generalisation of the CTM for the chiral Potts model.

The direct generalisations of the superintegrable chiral Potts model obtained from $\mathcal{A}_n(\alpha)$ lead to a new class of
superintegrable models. It would be interesting to find explicit examples of physical models which can be represented in this way. The presence of an 
Onsager algebra leads to some universal physical information, e.g. an Ising-like spectrum of the corresponding Hamiltonian \cite{Davies}.
A better knowledge of superintegrable models may yield more information about which features are
universal from the Onsager algebra. We have found that the isotropic XY spin chain is an example of this form, but additional examples would be valuable.

Finally, we note that two copies of TL algebras have previously appeared in the context of Lorentz lattice gases \cite{Martins:1998}. In that particular work, however, the two algebras are mutually commuting, leading to a rather different structure compared to our algebras $\mathcal{A}_n(\alpha)$.

\section*{Acknowledgements}
We would like to thank prof. J. H. H. Perk for pointing out references.
J.F. is supported in parts by NSFC grant No.~10775067  as well as   
Research Links Programme of the Swedish Research Council under contract No.~348-2008-6049.
The research by T.M.  is mainly supported by the Swedish Science Research Council, but also partly by the G\"oran Gustafsson 
foundation.

\begin{appendix}
\section{Useful identities in \texorpdfstring{$\mathcal{A}_n(\alpha)$}{An(a)}}\label{app:ids}
From the quadratic relations \eqref{rel3} and \eqref{rel4} one derives
\begin{equation}
\begin{split}
	& f_ie_{i\pm1}e_i = \alpha^{\mp 1} (e_{i\pm 1}e_i+\alpha^{\mp 2}f_{i \pm 1}e_i -  \alpha^{\mp 1} e_i)\\
	& e_ie_{i\pm 1}f_i=\alpha^{\pm 1}( e_{i\pm 1}f_i+\alpha^{\pm 2} f_{i\pm1}f_i - \alpha^{\pm 1} f_i)\\
	& f_if_{i \pm1}e_i = \alpha^{\pm 1}(\alpha^{\pm 2} e_{i\pm 1}e_i+f_{i\pm  1}e_i-\alpha^{\pm 1} e_i)\\
	& e_if_{i\pm 1}f_i =\alpha^{\mp 1}(\alpha^{\mp 2} e_{i\pm  1}f_i+f_{i\pm 1}f_i- \alpha^{\mp 1}f_i),
\end{split}\label{cubic3_med_A}
\end{equation}
which are equivalent with 
\begin{equation}
\begin{split}
	& f_ie_{i\pm 1}e_i  = \alpha^{\mp 1}(f_ie_{i\pm 1}+\alpha^{\mp 2}f_if_{i\pm 1} - \alpha^{\mp 1}f_i)\\
        & e_ie_{i\pm 1}f_i=\alpha^{\pm 1}( e_{i}e_{i\pm 1}+\alpha^{\pm 2} e_{i}f_{i\pm1} - \alpha^{\pm} e_i)\\
	& f_if_{i \pm 1}e_i = \alpha^{\pm 1}(\alpha^{\pm 2} f_{i}e_{i\pm 1}+f_{i}f_{i\pm 1}-\alpha^{\pm 1} f_i)\\
	& e_if_{i\pm 1}f_i =\alpha^{\mp 1}(\alpha^{\mp 2} e_{i}e_{i\pm 1}+e_{i}f_{i\pm 1}-\alpha^{\mp 1}e_i).
\end{split}\label{cubic4_med_A}
\end{equation}
Using these cubic relations it is straightforward to derive other useful relations.
The following two relations are used in the calculation of the first commuting charge.
\begin{equation}
\begin{split}
&e_i e_{i\mp 1} f_i+f_i f_{i\mp 1}e_i+f_ie_{i\mp 1}e_i+e_i f_{i\mp 1}f_i=(\alpha^{-2}+\alpha^2)\left(
\frac{1}{4}(\alpha+\alpha^{-1})\{e_{i\mp 1}+f_{i\mp 1},e_i+f_i\} \right. \\
&\left. \mp \frac{1}{4}(\alpha-\alpha^{-1})[f_{i\mp 1}-e_{i\mp 1},f_i-e_i]-(e_i+f_i) \right)=\mbox{\{only valid when eq.(31) is valid\}}=\\
&(\alpha+\alpha^{-1})\{e_{i\mp 1}+f_{i\mp 1},e_i+f_i\}-(\alpha^{-2}+\alpha^2)(e_i+f_i)
-(k_3 (e_i+f_i+e_{i\mp 1}+f_{i\mp 1})+k_4 )/4
\end{split}
\end{equation}
\begin{equation}
\label{Cubic_CC_Rel2}
\begin{split}
&e_{i} e_{i\mp 1} f_i-f_i f_{i\mp 1}e_i+f_ie_{i\mp 1}e_i-e_i f_{i\mp 1}f_i=\\
&\frac{1}{4}(\alpha^{3}+\alpha^{-3}-\alpha-\alpha^{-1})
\{f_{i\mp 1}- e_{i\mp 1},f_{i} + e_i\}
\mp \frac{1}{4}((\alpha^{3}-\alpha^{-3}+\alpha-\alpha^{-1}))[f_{i\mp 1} + e_{i\mp 1},f_{i} - e_i]= \\
& (\alpha-\alpha^{-1})(\mp [f_{i\mp 1} + e_{i\mp 1},f_{i} - e_i]+  (\alpha-\alpha^{-1})(f_{i\mp 1}-e_{i\mp 1}))
\end{split}
\end{equation}
The following relation is needed in the Onsager computation.
\begin{equation}
\label{Cubic_OA_Rel}
\begin{split}
&e_{i} e_{i\mp1} f_i-f_i e_{i\mp 1}e_i+f_if_{i\mp 1}e_i-e_i f_{i\mp 1}f_i=\\
&(\alpha^2-\alpha^{-2})\left(\mp
\frac{1}{4}(\alpha+\alpha^{-1})
\{e_{i\mp 1}+f_{i\mp 1},e_i+f_{i}\} 
+\frac{1}{4}(\alpha-\alpha^{-1}))[f_{i\mp 1}-e_{i\mp 1},f_{i}-e_i]
\pm (e_i+f_i) \right)\\
&=\mbox{\{only valid when eq.(31) is valid\}}\\
&=-(\alpha^2-\alpha^{-2})(\frac{(\alpha-\alpha^{-1})}{(\alpha^2+\alpha^{-2}-4)} [f_{i\mp1}-e_{i\mp1},f_{i}-e_{i}]
\mp \frac{(\alpha+\alpha^{-1})^2}{(\alpha^2+\alpha^{-2}-4)}(f_{i+1}+e_{i+1}+f_{i}+e_{i})\mp(e_i+f_i)+\mp k)
\end{split}
\end{equation}
In the last expression, $k$ is a numerical constant. 
\section{Details of boost operator calculations}\label{app:boost}
We start with the Hamiltonian
\begin{equation}
\begin{split}
&H =\sum_i  -kf(\bar{\phi},\phi)\cos {\phi/3} (e_{2i-1}+f_{2i-1})+f(\bar{\phi},\phi)\sin{\phi/3} (e_{2i-1}- f_{2i-1})\\
& -k f(\phi,\bar{\phi}) \cos {\bar{\phi}/3}(e_{2i}+f_{2i})+f(\phi,\bar{\phi})\sin{\bar{\phi}/3}(e_{2i}-f_{2i}).
\end{split}
\end{equation}
From the equation
\begin{equation}
\beta \frac{\partial H}{\partial \phi}+\bar{\beta}\frac{\partial H}{\partial \bar{\phi}}=cH+Q_E
\end{equation}
($c$ is a free parameter we get from the freedom to add something proportional to the Hamiltonian to $Q_E$)
we get the following equations:

\be
\begin{split}
&d_3- c k f(\phi,\bar{\phi})\cos \bar{\phi/3} =
(\frac{\bar{\beta}}{3}f(\phi,\bar{\phi} ) k \sin \bar{\phi}/3- (\beta\partial_{\phi}f(\phi,\bar{\phi})+\bar{\beta}\partial_{\bar{\phi}}f(\phi,\bar{\phi}))k\cos \bar{\phi}/3)\\
&d_4 + c f(\phi,\bar{\phi})\sin \bar{\phi}/3 =
(\frac{\bar{\beta}}{3}f(\phi,\bar{\phi})\cos \bar{\phi}/3+ (\beta \partial_{\phi}f(\phi,\bar{\phi})+\bar{\beta}\partial_{\bar{\phi}}f(\phi,\bar{\phi}))\sin \bar{\phi}/3)\\
&d_1- c k f(\bar{\phi},\phi)\cos \phi/3 =(\frac{\beta}{3} f(\bar{\phi},\phi) k \sin \phi/3 - (\beta \partial_{\phi}f(\bar{\phi},\phi)+\bar{\beta}\partial_{\bar{\phi}}f(\bar{\phi},\phi))
k\cos \phi/3)\\
&d_2 + c f(\bar{\phi},\phi)\sin \phi/3 =
( \frac{\beta}{3} f(\bar{\phi},\phi)\cos \phi/3 +(\beta \partial_{\phi}f(\bar{\phi},\phi)+\bar{\beta}\partial_{\bar{\phi}}f(\bar{\phi},\phi))\sin \phi/3 ).
\end{split}
\ee
Here
\be\label{sol2}
\begin{split}
& d_1=2 f(\phi,\bar{\phi})f(\bar{\phi},\phi) \sin {\phi/3}\sin \bar{\phi}/3 (\alpha-\alpha^{-1})C_2
\\
& d_2=- \frac1{k}f(\bar{\phi},\phi)\frac{\sin(\bar{\phi}/3)}{\cos(\bar{\phi}/3)} (\alpha-\alpha^{-1})
\left( f(\bar{\phi},\phi)(C_1\cos^2({\phi/3})+ 2 C_2-3)+2 f(\phi,\bar{\phi})k^2\cos(\phi/3)\cos(\bar{\phi}/3)\right)\\
&  d_3=2 f(\phi,\bar{\phi})f(\bar{\phi},\phi) \sin \bar{\phi}/3\sin {\phi/3}(\alpha-\alpha^{-1})C_2
 \\
& d_4=-\frac1{k} f(\phi,\bar{\phi})\frac{\sin(\phi/3)}{\cos(\phi/3)} (\alpha-\alpha^{-1})
\left(f(\phi,\bar{\phi})(C_1 \cos^2(\bar{\phi}/3)+2C_2-3)+
2 f(\bar{\phi},\phi)k^2\cos(\phi/3)\cos(\bar{\phi}/3)\right),\\
\end{split}
\ee 
where $C_1$ and $C_2$ are given in equation \eqref{Cparameters}. 
Note that we have here chosen $\tilde{\lambda_1}=f(\bar{\phi},\phi)$ and $\tilde{\lambda}_2=f(\phi,\bar{\phi})$, just as
in the Hamiltonian \eqref{kparam}, in contrast to the general expression of \eqref{genHam}.
We get four equations with four parameters ($c$, $\beta$, $\bar{\beta}$ and $k$).
Combining the first and second equation gives:
\be
\label{beta}
\begin{split}
&d_1 \sin \phi/3+d_2 k \cos \phi/3 =
\frac{\beta}{3} f(\bar{\phi},\phi) k \\
&d_1 \cos \phi/3-d_2 k \sin \phi/3-  c k f(\bar{\phi},\phi) =-
(\beta\partial_{\phi}f(\bar{\phi},\phi)+\bar{\beta}\partial_{\bar{\phi}}f(\bar{\phi},\phi))k.
\end{split}
\ee
Likewise for the last two rows:
\be
\label{barbeta}
\begin{split}
&d_3 \sin \bar{\phi}/3+d_4 k \cos \bar{\phi}/3 =
\frac{\bar{\beta}}{3}f(\phi,\bar{\phi}) k \\
&d_3 \cos \bar{\phi}/3-d_4 k \sin \bar{\phi}/3-  c k f(\phi,\bar{\phi}) =-
(\beta\partial_{\phi}f(\phi,\bar{\phi})+\bar{\beta}\partial_{\bar{\phi}}f(\phi,\bar{\phi}))k.
\end{split}
\ee

We use equations \eqref{barbeta} and \eqref{beta} to solve for $\bar{\beta}$ and $\beta$.
The other two equations can be combined in such a way that we get rid of the $c$ dependence, and we obtain an equation 
for $k$: 
\begin{equation}
\begin{split}
& f(\phi,\bar{\phi})f(\bar{\phi},\phi)(d_{3ex}\cos \bar{\phi}/3-d_{4kex}\sin \bar{\phi}/3) -
(d_{3ex}\sin \bar{\phi}/3+ d_{4kex}\cos \bar{\phi}/3 )\left(f(\phi,\bar{\phi})\partial_{\bar{\phi}}f(\bar{\phi},\phi)-
f(\bar{\phi},\phi)\partial_{\bar{\phi}} f(\phi, \bar{\phi})\right)-\\
&f(\phi,\bar{\phi})f(\bar{\phi},\phi)(d_{1ex}\cos {\phi}/3-d_{2kex}\sin {\phi/3}) +
(d_{1ex}\sin {\phi/3}+ d_{2kex}\cos {\phi/3} )\left(f(\bar{\phi},\phi)\partial_{\bar{\phi}}f(\phi,\bar{\phi})-
f(\phi, \bar{\phi} )\partial_{\bar{\phi}} f(\bar{\phi},\phi)\right)=0,
\end{split}
\end{equation}
where 
\be
\label{eq:dex}
d_{1ex}=\frac{d_1}{f(\bar{\phi},\phi)} \qquad d_{2kex}=k\frac{d_2}{f(\bar{\phi},\phi)} 
\qquad d_{3ex}=\frac{d_1}{f(\phi,\bar{\phi})} \qquad d_{4kex}=k\frac{d_4}{f(\phi,\bar{\phi})}.
\ee
Note that these terms do not contain $k$ in the denominator.
\section{Details of Onsager calculations}\label{app:onsager}
Here we have collected some important details used in verifying the Dolan-Grady conditions. Let us
consider the left hand side of equation \eqref{Dolan_Grady} (summation over the index $i$ is suppressed).
\begin{equation}
\begin{split}
[A_0[A_0[A_0,A_1]]]&=6[e_{2i}\, , e_{2i+2}e_{2i+1}f_{2i+2}+f_{2i+2}e_{2i+1}e_{2i+2}-e_{2i+2}f_{2i+1}f_{2i+2}-f_{2i+2}f_{2i+1}e_{2i+2}]\\
&+6[e_{2i+2}\, , e_{2i}e_{2i+1}f_{2i}+f_{2i}e_{2i+1}e_{2i}-e_{2i}f_{2i+1}f_{2i}-f_{2i}f_{2i+1}e_{2i}]\\
&+6\gamma (e_{2i+2}e_{2i+1}f_{2i+2}-f_{2i+2}e_{2i+1}e_{2i+2}-e_{2i+2}f_{2i+1}f_{2i+2}+f_{2i+2}f_{2i+1}e_{2i+2})\\
&+6\gamma (e_{2i}e_{2i+1}f_{2i}-f_{2i}e_{2i+1}e_{2i}-e_{2i}f_{2i+1}f_{2i}+f_{2i}f_{2i+1}e_{2i})\\
&+6\gamma(f_{2i+2}f_{2i+1}e_{2i}-e_{2i+2}f_{2i+1}f_{2i}+f_{2i}f_{2i+1}e_{2i+2}-e_{2i}f_{2i+1}f_{2i+2})\\
&+6\gamma(e_{2i+2}e_{2i+1}f_{2i}-f_{2i+2}e_{2i+1}e_{2i}+e_{2i}e_{2i+1}f_{2i+2}-f_{2i}e_{2i+1}e_{2i+2})\\
&+6\gamma(e_{2i}e_{2i+2}e_{2i+1}-f_{2i}f_{2i+2}e_{2i+1}+e_{2i+1}f_{2i+2}f_{2i}-e_{2i+1}e_{2i+2}e_{2i})\\
&+6\gamma(f_{2i}f_{2i+2}f_{2i+1}+f_{2i+1}e_{2i+2}e_{2i}-f_{2i+1}f_{2i+2}f_{2i}-e_{2i}e_{2i+2}f_{2i+1})\\
&+\gamma^3 [A_0,A_1]
\end{split}
\end{equation}
The first two lines can be rewritten using equation \eqref{Cubic_CC_Rel2}:
\begin{equation}
\begin{split}
&[e_{2i}\, , e_{2i+2}e_{2i+1}f_{2i+2}+f_{2i+2}e_{2i+1}e_{2i+2}-e_{2i+2}f_{2i+1}f_{2i+2}-f_{2i+2}f_{2i+1}e_{2i+2}]\\
&+[e_{2i+2}, e_{2i}e_{2i+1}f_{2i}+f_{2i}e_{2i+1}e_{2i}-e_{2i}f_{2i+1}f_{2i}-f_{2i}f_{2i+1}e_{2i}]\\
&=-(\alpha-\alpha^{-1})^2[e_{2i}-f_{2i}+e_{2i+2}-f_{2i+2},e_{2i+1}-f_{2i+1}].
\end{split}
\end{equation}
Lines three and four can be written as (using \eqref{Cubic_OA_Rel}):
\begin{equation}
\begin{split}
&+\gamma (e_{2i+2}e_{2i+1}f_{2i+2}-f_{2i+2}e_{2i+1}e_{2i+2}-e_{2i+2}f_{2i+1}f_{2i+2}+f_{2i+2}f_{2i+1}e_{2i+2})\\
&+\gamma (e_{2i}e_{2i+1}f_{2i}-f_{2i}e_{2i+1}e_{2i}-e_{2i}f_{2i+1}f_{2i}+f_{2i}f_{2i+1}e_{2i})\\
&=(\frac{(\alpha^2-\alpha^{-2})^2}{(\alpha^2+\alpha^{-2}-4)} [f_{i}-e_{i},f_{i-1}-e_{i-1}+f_{i+1}-e_{i+1}].
\end{split}
\end{equation}
The above equality is only valid up to terms disappearing from periodicity in a periodic chain. Note in equation \eqref{Cubic_OA_Rel}, however,
that for the special case $\alpha^2=\alpha^{-2}$ one does not need periodicity.
Lines five to eight are a bit more complicated.
The seventh line can be rewritten in two different ways,  either as:
\begin{equation}
\begin{split}
&e_{2i}e_{2i+2}e_{2i+1}-f_{2i}f_{2i+2}e_{2i+1}-e_{2i+1}e_{2i+2}e_{2i}+e_{2i+1}f_{2i+2}f_{2i}=\\
&-\alpha^{-2}(f_{2i}f_{2i+1}e_{2i+2}-e_{2i}f_{2i+1}f_{2i+2}+e_{2i+2}f_{2i+1}f_{2i}-f_{2i+2}f_{2i+1}e_{2i})\\
&+\alpha^{-1}(e_{2i}(e_{2i+1}-f_{2i+1})-(e_{2i+1}-f_{2i+1})e_{2i+2}-f_{2i+2}(e_{2i+1}-f_{2i+1})+(e_{2i+1}-f_{2i+1})f_{2i}),
\end{split}
\end{equation}
or as:
\begin{equation}
\begin{split}
&e_{2i}e_{2i+2}e_{2i+1}-f_{2i}f_{2i+2}e_{2i+1}-e_{2i+1}e_{2i+2}e_{2i}+e_{2i+1}f_{2i+2}f_{2i}=\\
&-\alpha^2(f_{2i+2}f_{2i+1}e_{2i}-e_{2i+2}f_{2i+1}f_{2i}+f_{2i}f_{2i+1}e_{2i+2}-e_{2i}f_{2i+1}f_{2i+2})\\
&-\alpha( f_{2i}(e_{2i+1}-e_{2i+2}-f_{2i+1}+f_{2i+2})+e_{2i+2}(-e_{2i}-e_{2i+1}+f_{2i}+f_{2i+1})).
\end{split}
\end{equation}
Choosing an arbitrary combination of these it follows that this expression can be written as:
\begin{equation}
\begin{split}
&e_{2i}e_{2i+2}e_{2i+1}-f_{2i}f_{2i+2}e_{2i+1}-e_{2i+1}e_{2i+2}e_{2i}+e_{2i+1}f_{2i+2}f_{2i}=\\
&-(x\alpha^2+(1-x)\alpha^{-2})(f_{2i+2}f_{2i+1}e_{2i}-e_{2i+2}f_{2i+1}f_{2i}+f_{2i}f_{2i+1}e_{2i+2}-e_{2i}f_{2i+1}f_{2i+2})\\
&-x\alpha( f_{2i}(e_{2i+1}-f_{2i+1}+f_{2i+2})+e_{2i+2}(-e_{2i}-e_{2i+1}+f_{2i+1}))\\
&+(1-x)\alpha^{-1}(e_{2i}(e_{2i+1}-f_{2i+1})-(e_{2i+1}-f_{2i+1})e_{2i+2}-f_{2i+2}(e_{2i+1}-f_{2i+1})+(e_{2i+1}-f_{2i+1})f_{2i}),
\end{split}
\end{equation}
where $x$ is a free parameter.
Finally this allows us to write lines five to eight as:
\begin{equation}
\begin{split}
&6\gamma(f_{2i+2}f_{2i+1}e_{2i}-e_{2i+2}f_{2i+1}f_{2i}+f_{2i}f_{2i+1}e_{2i+2}-e_{2i}f_{2i+1}f_{2i+2})\\
&+6\gamma(e_{2i+2}e_{2i+1}f_{2i}-f_{2i+2}e_{2i+1}e_{2i}+e_{2i}e_{2i+1}f_{2i+2}-f_{2i}e_{2i+1}e_{2i+2})\\
&+6\gamma(e_{2i}e_{2i+2}e_{2i+1}-f_{2i}f_{2i+2}e_{2i+1}+e_{2i+1}f_{2i+2}f_{2i}-e_{2i+1}e_{2i+2}e_{2i})\\
&+6\gamma(f_{2i}f_{2i+2}f_{2i+1}+f_{2i+1}e_{2i+2}e_{2i}-f_{2i+1}f_{2i+2}f_{2i}-e_{2i}e_{2i+2}f_{2i+1})\\
&=6\gamma(1-(x\alpha^2+(1-x)\alpha^{-2})(\mbox{cubic terms})
+6\gamma(x\alpha+(1-x)\alpha^{-1}) [A_0,A_1].
\end{split}
\end{equation}
Choosing $x$ (when $\alpha^2\neq \alpha^{-2}$) according to
\begin{equation}
x=\frac{\alpha^{-2}-1}{\alpha^{-2}-\alpha^2} \quad \Rightarrow\quad 6\gamma(-x\alpha+(1-x)\alpha^{-1})=12,
\end{equation}
 all cubic terms disappear, resulting in
\begin{equation}
[A_0[A_0[A_0,A_1]]]=(12-6(\alpha-\alpha^{-1})^2+6\frac{(\alpha^2-\alpha^{-2})^2}{\alpha^2+\alpha^{-2}-4}+\gamma^2 ) [A_0,A_1].
\end{equation}
or maybe better
\begin{equation}
\label{Onsager_formula}
[A_0[A_0[A_0,A_1]]]=(12+36\frac{(\alpha-\alpha^{-1})^2}{\alpha^2+\alpha^{-2}-4}+\gamma^2 ) [A_0,A_1].
\end{equation}
Note that the cubic terms automatically disappears for $\alpha^2= \alpha^{-2}$, and one also gets 
a factor of $12$ in front of $ [A_0,A_1]$ from line five to eight. The case $\alpha=\pm i$ needs special treatment, since then the formula \eqref{Onsager_formula}
is not applicable. Only the first two rows then give the contribution $24$ to the factor in front of $ [A_0,A_1]$

\section{A $9$-dimensional representation of $\mathcal{A}_2(5/2)$}
\label{small}
Using GAP we obtain the following nine dimensional representation of the
algebra $\mathcal{A}_2(5/2)=\mathcal{A}_3^o(5/2)$.
\begin{equation}
e_1=\left(\begin{array}{ccccccccc}  0 &1& 0& 0& 0& 0& 0& 0& 0 \\ 0& 5/2& 0& 0 &0 &0 &0 &0 &0 \\ 0 &0 &0 &0 &0 &0 &0 &0 &0 \\ 17/12& -1/6& -2/3& -1/6& -2/3& 1/15& 4/15& 4/15& 19/60 \\ -17/3& 8/3& 2/3& 2/3& 8/3& -4/15& -16/15 &44/15& -4/15 \\ 0& 1& 0 &0 &0 &0 &0 &0 &0 \\ 0& 1& 0& 0& 0& 0& 0& 0& 0 \\ 0& 0& -1/4& 0 &0 &0 &0 &1/2& 1/8\\ 0 &0 &-4 &0 &0 &0 &0 &8 &2 \end{array}
\right)
\end{equation}
\begin{equation}
f_1=\left(\begin{array}{ccccccccc} 0 &0 &1 &0 &0 &0 &0 &0 &0 \\ 0& 0 &0 &0 &0 &0 &0 &0 &0 \\ 0 &0 &5/2& 0 &0& 0& 0& 0& 0\\ -17/3 &2/3& 8/3 &8/3& 2/3& -4/15& 44/15& -16/15& -4/15 \\ 17/12& -2/3& -1/6 &-2/3 &-1/6 &19/60& 4/15& 4/15& 1/15 \\ 0& -4 &0 &0 &0& 2 &8 &0 &0 \\ 0& -1/4& 0 &0 &0 &1/8& 1/2& 0& 0\\ 0& 0 &1 &0 &0 &0 &0 &0 &0\\ 0& 0& 1 &0 &0 &0& 0& 0& 0\end{array}
\right)
\end{equation}
\begin{equation}
e_2=\left(\begin{array}{ccccccccc} 0 &0& 0& 1& 0& 0& 0& 0& 0\\ 0& 0& 0 &0 &0& 1& 0& 0& 0\\ 0& 0 &0 &0& 0 &0 &0 &1& 0\\ 0& 0& 0& 5/2& 0& 0& 0& 0& 0\\ 0& 0 &0 &0 &0 &0 &0 &0& 0\\ 0& 0& 0& 0& 0& 5/2& 0 &0 &0\\ 0& 0 &0 &0 &0 &0 &0 &0& 0\\ 0& 0 &0 &0 &0 &0 &0 &5/2& 0\\ 0& 0& 0& 0& 0& 0& 0 &0 &0 
          \end{array}\right)
\quad
f_2= \left(\begin{array}{ccccccccc} 0 &0 &0 &0 &1 &0 &0 &0 &0\\ 0& 0& 0& 0& 0& 0& 1& 0& 0\\ 0 &0 &0 &0 &0 &0 &0 &0 &1\\ 0& 0& 0& 0& 0& 0& 0& 0& 0\\ 0& 0& 0& 0& 5/2& 0 &0 &0 &0\\ 0& 0& 0& 0& 0& 0& 0& 0& 0\\ 0& 0& 0& 0& 0& 0& 5/2& 0& 0\\ 0& 0& 0& 0& 0& 0& 0& 0& 0\\ 0& 0& 0 &0 &0 &0 &0 &0 &5/2 
            \end{array}\right)
\end{equation}

\end{appendix}

\bibliographystyle{hunsrt}

\end{document}